\documentclass[10pt]{article}
\usepackage{graphicx}

\def\Title#1{\begin{center} {\Large #1 } \end{center}}
\def\Author#1{\begin{center}{ \sc #1} \end{center}}
\def\Address#1{\begin{center}{ \it #1} \end{center}}

\newcommand\pubblock{\rightline{\begin{tabular}{l} Proceedings of the Fifth Annual LHCP\\ \pubnumber\\
         \pubdate  \end{tabular}}}

\newenvironment{Abstract}{\begin{quotation} \begin{center} 
             \large ABSTRACT \end{center}\bigskip 
      \begin{center}\begin{large}}{\end{large}\end{center} \end{quotation}}

\newenvironment{Presented}{\begin{quotation} \begin{center} 
             PRESENTED AT\end{center}\bigskip 
      \begin{center}\begin{large}}{\end{large}\end{center} \end{quotation}}

\def\beq{\begin{equation}}
\def\eeq#1{\label{#1}\end{equation}}
\def\eeqn{\end{equation}}

\def\beqa{\begin{eqnarray}}
\def\eeqa#1{\label{#1}\end{eqnarray}}
\def\eeqan{\end{eqnarray}}

\let\bar=\overbar

\def\Dslash{\not{\hbox{\kern-4pt $D$}}}
\def\dslash{\not{\hbox{\kern-2pt $\del$}}}

\def\msb{{\bar{\ssstyle M \kern -1pt S}}}

\usepackage{color}
\usepackage{amsmath}
\usepackage{xspace}
\usepackage{esvect}
\RequirePackage{lineno}

\def\CP{{\ensuremath{C\!P}}\xspace}
\def\phis{{\ensuremath{\phi_s}}\xspace}
\def\Lb{{\ensuremath{\varLambda^{0}_{b}}}\xspace}
\def\Bs{\ensuremath{B_{s}^{0}}\xspace}
\def\jpsi{{\ensuremath{{J\mskip -3mu/\mskip -2mu\psi\mskip 2mu}}}\xspace}
\def\Bspsiphi{{\ensuremath{\Bs\to\psi(2S)\phi}}\xspace}
\def\BsjpsiKK{{\ensuremath{\Bs\to\jpsi K^{+}K^{-}}}\xspace}

\newcommand\pubnumber{ LHCB-PROC-2017-032 }
\newcommand\pubdate{September 1, 2017}

\def\affiliation{
On behalf of the LHCb Collaboration, \\
Center for High Energy Physics, \\
Tsinghua University, Beijing, 100084, China}

\begin{document}

\large
\begin{titlepage}
\pubblock

\vfill
\Title{ Mixing and \CP violation results in $b$-hadron decays at LHCb }
\vfill

\Author{ XUESONG LIU  }
\Address{\affiliation}
\vfill
\begin{Abstract}

Measurements of mixing and \CP violation in $b$-hadron decays are great probes to search for physics beyond the Standard Model.
A selection of recent results based upon $3.0\ \mathrm{fb}^{-1}$ of LHCb data are presented.  
These are the first measurement of the \CP violating phase \phis using \BsjpsiKK in the mass region above the $\phi(1020)$ resonance and \Bspsiphi decays, the determination of \Bs  lifetime using $\Bs\to D^{(*)-}\mu^{+}\nu_{\mu}$ decays, the updated measurements of the CKM angle $\gamma$ with $B^{\pm}\to D K^{*\pm}$ and $\Bs\to D_{s}^{\mp}K^{\pm}$ decay modes and the \CP violation searches in $\Lb\to p\pi^{-}\pi^{+}\pi^{-}$ and $\Lb\to pK^{-}\mu^{+}\mu^{-}$ decays.

\end{Abstract}
\vfill

\begin{Presented}
The Fifth Annual Conference\\
 on Large Hadron Collider Physics \\
Shanghai Jiao Tong University, Shanghai, China\\ 
May 15-20, 2017
\end{Presented}
\vfill
\end{titlepage}
\def\thefootnote{\fnsymbol{footnote}}
\setcounter{footnote}{0}

\normalsize 

\section{Measurements of the CP violating phase \phis}
The interference of the decay  amplitudes and the \Bs-$\overline{B}{}^{0}_{s}$ mixing amplitudes can give rise to a \CP violating phase
\begin{equation}
\phis=\phi_{M}-2\phi_{D},
\end{equation}
where $\phi_{M}$ is the \Bs-$\overline{B}{}^{0}_{s}$ mixing phase and $\phi_{D}$ is the weak phase between the \Bs and $\overline{B}{}^{0}_{s}$ decay amplitudes. Measurement of the \phis is obtained through the time-dependent \CP asymmetry
\begin{equation}
A_{\CP}(t)=\frac{\Gamma(\overline{B}{}^{0}_{s}(t)-f)-\Gamma(\Bs(t)-f)}{\Gamma(\overline{B}{}^{0}_{s}(t)-f)+\Gamma(\Bs(t)-f)}=\frac{C_{f}\mathrm{cos}(\Delta mt)-S_{f}\mathrm{sin}(\Delta mt)}{\mathrm{cosh}(\Delta \Gamma t/2)+A^{\Delta \Gamma}\mathrm{sinh}(\Gamma t/2)},
\end{equation}
where $C_{f}$ is related to \CP violation in the decay and $S_{f}$ represents \CP violation in the interference. 
Neglecting penguin pollution and assuming no Beyond Standard Model (BSM) contribution, the phase $\phis$ is predicted to be $\phis=-2\beta_{s}=0.0365^{+0.0013}_{-0.0012}\ \mathrm{rad}$~\cite{Charles:2015gya}.
The world average result by HFLAV, $\phis=-0.021\pm0.031\ \mathrm{rad}$~\cite{Amhis:2016xyh}, is dominated by LHCb measurements and is compatible with Standard Model (SM) prediction.

\subsection*{Measurement of \phis with \BsjpsiKK in the mass region above the $\phi(1020)$ }
A study performed by LHCb, using a data sample corresponding to $1\ \mathrm{fb^{-1}}$ showes a rich resonance spectrum in the $K^{+}K^{-}$ mass distribution~\cite{Aaij:2013orb}, indicating that the signficant contribution from the $f^{'}_{2}(1525)$ resonance and nonresonant S-wave amplitudes can be used for further studies of \CP violation~\cite{Aaij:2011ac}. 
In this analysis, the \CP observables are determined as functions of the \Bs decay time in a four-dimensional phase space including the three helicity angles characterizing the decay and the invariant mass of the two kaons, $m_{KK}$. 
With respect to the previous publication focusing on the $\jpsi\phi(1020)$ decays~\cite{Aaij:2014zsa}, the $\BsjpsiKK$ sample is separated into two regions, having $m_{KK}$ above or below the $\phi(1020)$ resonance.
A simultaneous fit is performed for the two samples. 

\begin{figure}[htb]
\centering
\includegraphics[width=0.50\textwidth]{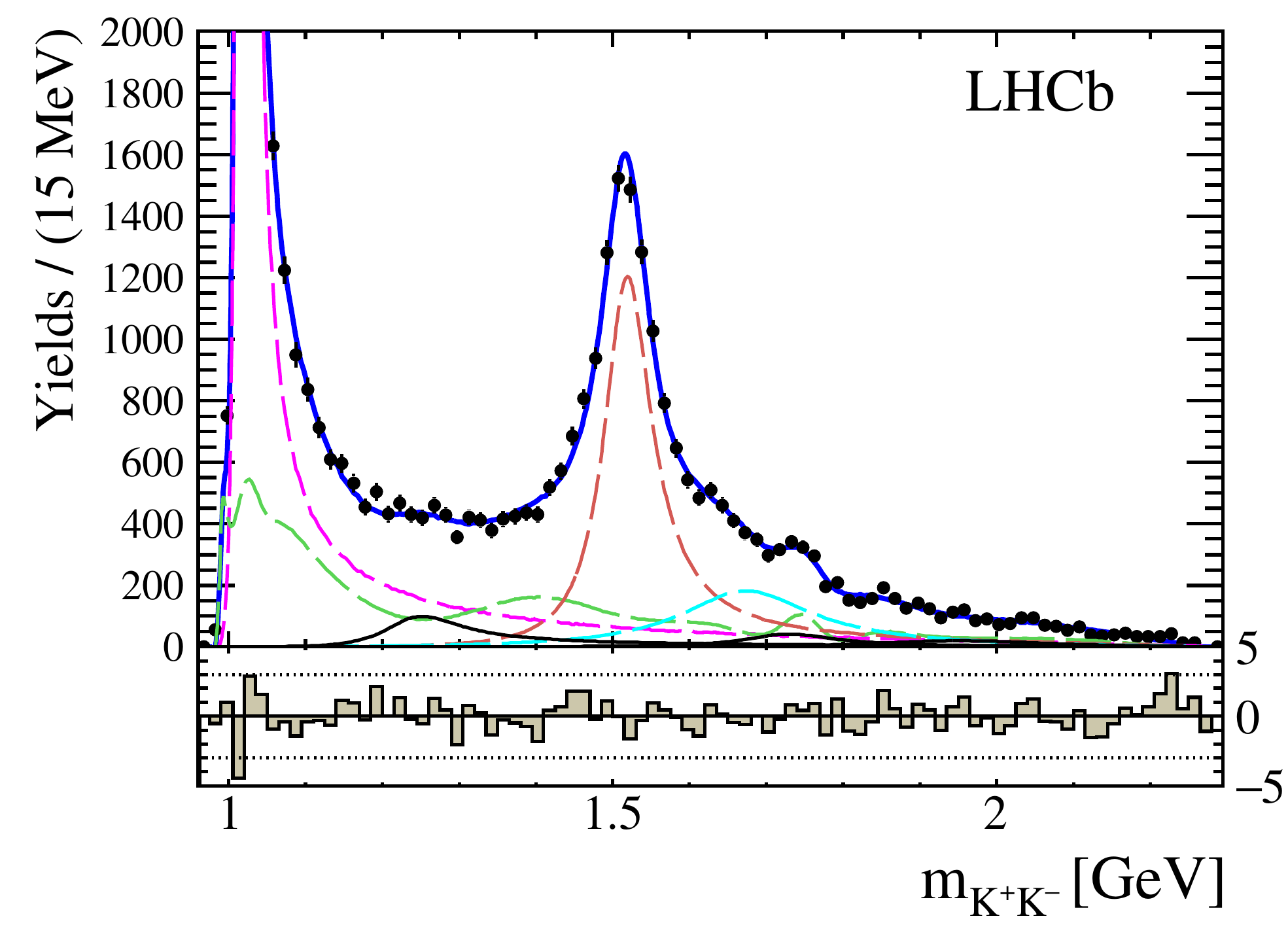}
\caption{Fit projection of $m_{KK}$. The resonance $\phi(1020)$, $f^{'}_{2}(1525)$, $\phi(1680)$ and S-wave components are shown by magenta, brown, cyan and green curves; the other $f_2$ resonances are described by black solid curves.}
\label{fig:figure4}
\end{figure}

All the resonanses found in the previous analysis are considered as contribution to the entire fit, except for the unconfirmed $f_2(1640)$ state. 
The fit projection of $m_{KK}$ is shown in Figure \ref{fig:figure4}~\cite{Aaij:2017zgz}.
The results are determined as
\begin{equation*}
\begin{array}{rl}
\phis =& 0.119\pm0.107\pm0.034\ \mathrm{rad},\\
\Delta\Gamma_s=&0.066\pm0.018\pm0.010\ \mathrm{ps}^{-1},\\
m_{f^{'}(1525)}=&1522.2\pm1.3\pm1.1,\ \mathrm{MeV},\\
\Gamma_{f^{'}(1525)}=&78.0\pm3.0\pm3.7,\ \mathrm{MeV},
\end{array}
\end{equation*}
where $\Delta\Gamma_s$ is the decay width difference between the light and the heavy mass eigenstates, $m_{f^{'}(1525)}$ and $\Gamma_{f^{'}(1525)}$ are the mass and width of $f^{'}(1525)$, the first uncertainty is statistical and the second is systematic.

The combination with previous results from \Bs decays in the $\phi(1020)$ region~\cite{Aaij:2014zsa} and $\jpsi \pi^{+}\pi^{-}$ final states~\cite{Aaij:2014dka} gives $\phis=0.001\pm0.037\ \mathrm{rad}$~\cite{Aaij:2017zgz}.

\subsection*{Measurement of \phis with \Bspsiphi}
The $\psi(2S)$ resonance state is for the first time used in the \phis measurement with $\Bs\to\phi(2S)(\mu^{+}\mu^{-})\phi(K^{+}K^{-})$ decay mode. 
This decay has a similar topology compared with the `golden mode' $\Bs\to \jpsi(\mu^{+}\mu^{-})\phi( K^{+}K^{-})$ ~\cite{Aaij:2014zsa}. 
Compared to the latter analysis, a multivariate selection is added to optimise the statistical sensitivity and the control mode $\Bs\to \phi(2S)K^*$ is used to calibrate the decay time acceptance.
In total, $4697\pm71$ signal candidates are found in this channel. 
An unbinned maximum likelihood fit to the four-dimensional distributions of the \Bspsiphi decay time and helicity angles is done to estimated the \CP violation phase \phis. The results, listed in Table \ref{tab:table1}~\cite{Aaij:2016ohx}, are compatible with previous LHCb measurements~\cite{Aaij:2014zsa, Aaij:2014dka}. 
\begin{table}[t]
\begin{center}
\begin{tabular}{l|c}  
Parameter&Value\\ \hline
$\Gamma_{s}$ [ $\mathrm{ps}^{-1}$ ]&$0.668\pm0.011\pm0.006$\\
$\Delta\Gamma_{s}$ [ $\mathrm{ps}^{-1}$ ]&$0.066^{+0.041}_{-0.044}\pm0.007$\\
$|A_{\perp}|^{2}$&$0.264^{+0.024}_{-0.023}\pm0.002$\\
$|A_{0} |^{2}$&$0.422\pm0.014\pm0.003$\\
$\delta_{\parallel}$ [$\mathrm{rad}$]&$3.67^{+0.13}_{-0.18}\pm0.03$\\
$\delta_{\perp}$ [$\mathrm{rad}$]&$3.29^{0.43}_{0.39}\pm0.04$\\
\phis [$\mathrm{rad}$]&$0.23^{+0.29}_{-0.28}\pm0.02$\\
$|\lambda|$&$1.045^{+0.069}_{-0.050}\pm0.007$\\
$F_{S}$&$0.061^{+0.026}_{-0.025}\pm0.007$\\
$\delta_{S}$ [$\mathrm{rad}$]&$0.03\pm0.14\pm0.02$\\ \hline
\end{tabular}
\caption{ Results of the maximum likelihood fit to the selected \Bspsiphi candidates. The parameters are defined in Ref.~\cite{Aaij:2016ohx}. The first uncertainty is statistical and the second is systematic. }
\label{tab:table1}
\end{center}
\end{table}

\section{Flavour-specific \Bs lifetime measurement with $\Bs\to D^{(*)-}_{s}\mu^{+}\nu_{\mu}$ }
The flavour-specific lifetime is defined as
\begin{equation}
\tau^{\mathrm{fs}}_{s}=\frac{1}{\Gamma_{s}}\left[\frac{1+(\Delta\Gamma_{s}/2\Gamma_{s})^{2}}{1-(\Delta\Gamma_{s}/2\Gamma_{s})^{2}}\right],
\end{equation}
where $\Delta\Gamma_{s}=\Gamma_{s,\mathrm{H}}-\Gamma_{s,\mathrm{L}}$ is the decay width difference between the light and the heavy mass eigenstates and $\Gamma_{s}=(\Gamma_{s,\mathrm{H}}+\Gamma_{s,\mathrm{L}})/2$ is the average width.

LHCb has reported a measurement of $\tau^{\mathrm{fs}_{s}}=1.535\pm0.015\pm0.014\ \mathrm{ps}$~\cite{Aaij:2014sua} using $\Bs\to D_{s}^{-}\pi^{+}$ decays. 
Semileptonic \Bs decays have a larger signal yield than in hadronic decays, such that can offer richer potential for precise measurement. 
In this analysis, the \Bs candidates are reconstructed with $D_{s}^{(*)-}\to K^{+}K^{-}\pi^{-}$ and muon candidates. 
The \Bs yields are determined with the `corrected-mass', defined as \mbox{$m_{\rm corr}=p_{\perp,D\mu}+\sqrt{m^{2}_{D\mu}+p^{2}_{\perp,D\mu}}$}. 
The lifetime is determined from the variation in the \Bs signal yield as a function of decay time, relative to that of $B^{0}$ decays that are reconstructed in the same final state. 
The result is $\tau^{\mathrm{fs}}_{s}=1.547\pm0.013(\mathrm{stat})\pm0.010(\mathrm{syst})\pm0.004(\tau_{B})\ \mathrm{ps}$~\cite{Aaij:2017vqj} and is consistent with and more precise than the current measured values~\cite{Aaij:2014sua,Abazov:2014rua}.

\section{Measurements of the CKM angle $\gamma$}
The angle $\gamma$ is the least well known angle in the CKM unitarity triangle.
It can be measured using tree-level processes based on the interference between $b\to c\bar{u}s$ and $b\to u\bar{c}s$ transitions and has negligible theoretical uncertainty~\cite{Brod:2013sga} in SM predictions.
Disagreement between such direct measurement of $\gamma$ and the value inferred from global CKM fits would indicate new physics beyond the SM. 
The world-average value of the $\gamma$ measurements with ($B\to DK$)-like decays is $\gamma=(74.0^{+5.8}_{-6.4})^{\circ}$~\cite{Amhis:2016xyh}.
The CKMFitter group quoted $\gamma=(66.9^{+0.9}_{-3.4})^{\circ}$~\cite{Charles:2004jd} from a global fit of CKM parameters.
And the latest LHCb combination yields $\gamma=(76.8^{+5.1}_{-5.7})^{\circ}$~\cite{Aaij:2016kjh,Aaij:2017conf}.
Exploiting the Run-II dataset with already used or even new decay modes will allow to reduce the uncertainty on the measured value of $\gamma$.

\subsection*{Measurement of the \CP observables using $B^{\pm}\to D K^{*\pm}$}
The \CP violation in $B^{\pm}\to D K^{*\pm}$ decays is sensitive to $\gamma$ at tree level.
Decays of the $D$ mesons to $ K^{-}\pi^{+},K^{+}K^{-},\pi^{+}\pi^{-},K^{+}\pi^{-},K^{-}\pi^{+}\pi^{-}\pi^{+},\pi^{+}\pi^{-}\pi^{+}\pi^{-},K^{+}\pi^{-}\pi^{+}\pi^{-}$ and the $K^{*}(892)^{\pm}$ with the $ K^{0}_{s}\pi^{\pm}$ final states are used in the reconstruction.
LHCb has reported the measurement of $\gamma$ with $B^{\pm}\to D K^{\pm}$ decays modes~\cite{Aaij:2016oso}. Compared with these channels, the branching fractions of $B^{\pm}\to D K^{*\pm}$ is of a similar magnitude, however the reconstruction efficiencies for the $K^{*\pm}$ decay are much lower. 

\begin{figure}[htb]
\centering
\includegraphics[width=\textwidth]{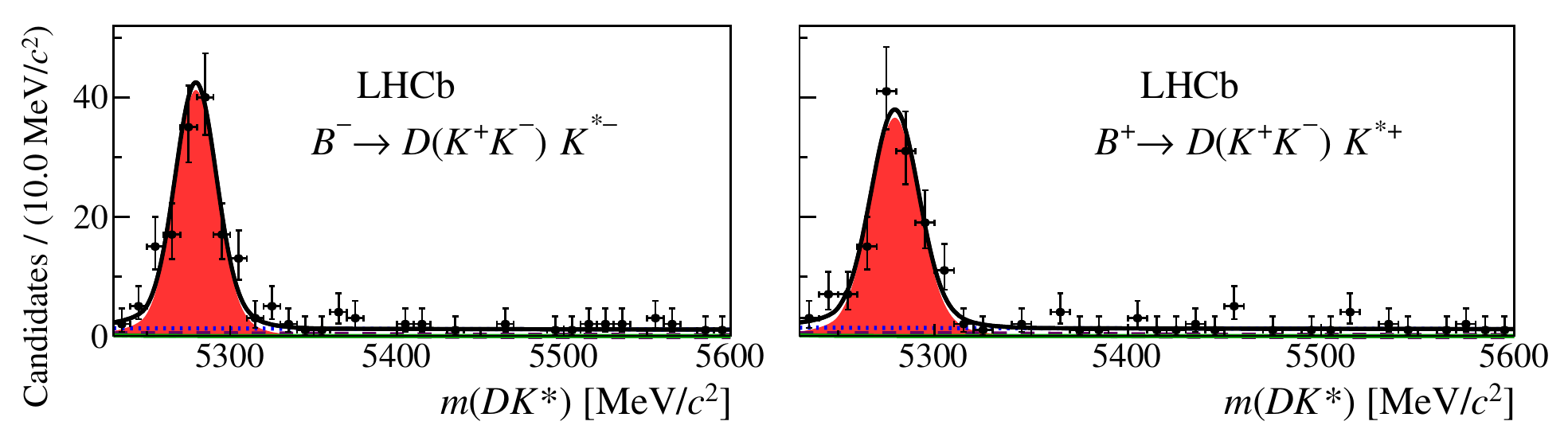}
\caption{ Mass distribution of the $B^{\pm}\to D(K^+K^-) K^{*\mp}$ modes using Run 1 and Run 2 data. The signal component is represented by the red shaded area, the combinatorial background by the dotted blue line and the partially reconstructed background by the solid green line.}
\label{fig:figure2}
\end{figure}

Several \CP asymmetries and ratios, corresponding to different final state of the $D^{0}$, were measured as the $\gamma$ sensitive observables.
The mass spectrum of the $B^{\pm}\to D(K^+K^-) K^{*\mp}$ decay modes are shown in Figure \ref{fig:figure2}~\cite{Aaij:2017paper}. A simultaneous fit is performed to all $D$ final state.

The measured \CP ovservable in the favoured mode,  $A_{K\pi}=-0.004\pm0.023(\mathrm{stat})\pm0.008(\mathrm{syst})$~\cite{Aaij:2017paper} , is consistent with zero. The combined results from the \CP-eigenstate $K^{+}K^{-}$ and $\pi^{+}\pi^{-}$ decay modes, $R_{\CP+}=1.18\pm0.08(\mathrm{stat})\pm0.01(\mathrm{syst}),\ A_{\CP+}=0.08\pm0.06(\mathrm{stat})\pm0.01(\mathrm{syst})$~\cite{Aaij:2017paper}, are consistent with and more precise than the previous measurements from BaBar~\cite{Aubert:2009cp}.

\subsection*{Measurement of the CKM angle $\gamma$ using $\Bs\to D^{\mp}_{s}K^{\pm}$}
Another approach to extract $\gamma$ is through the neutral $B$ meson decays and perform a time-depentment measurement of the \CP observables. 
In $\Bs\to D^{\mp}_{s}K^{\pm}$ decays, the sensitivity to the weak phase $\gamma-2\beta_{s}$ arises from the interference of \Bs and $\overline{B}{}^{0}_{s}$ decaying into the same final state : $D_{s}^{+}K^{-}$ or $D_{s}^{-}K^{+}$.
The \CP observables can be interpreted in terms of either $\gamma$ or $\beta_{s}$ by using an independent measurement of the other parameters as input and are given by
\begin{equation}
\begin{array}{rl}
C_{f}=\frac{1-r_{D_s K}^2 }{1+r_{D_s K}^2},\  A_{f}^{\Delta\Gamma}=\frac{-2r_{D_{s}K}\mathrm{cos}(\delta-(\gamma-2\beta_{s}))}{1+r^{2}_{D_s K}},&\ A_{\bar{f}}^{\Delta\Gamma}=\frac{-2r_{D_{s}K}\mathrm{cos}(\delta+(\gamma-2\beta_{s}))}{1+r^{2}_{D_s K}},\\
S_{f}=\frac{2r_{D_{s}K}\mathrm{sin}(\delta-(\gamma-2\beta_{s}))}{1+r^{2}_{D_s K}},&\ S_{\bar{f}}=\frac{-2r_{D_{s}K}\mathrm{sin}(\delta+(\gamma-2\beta_{s}))}{1+r^{2}_{D_s K}},\\
\end{array}
\end{equation}
where $r_{D_{s} K}\equiv |\lambda_{D_{s} K}|=|A(\overline{B}{}^{0}_{s}\to D_{s}^{-}K^{+})/A(\Bs\to D_{s}^{-}K^{+})|$ is the magnitude of the amplitude ratio, $\delta$ is the strong phase difference between the favoured and suppressed amlitudes. The intial flavour of the $\Bs$ mesons is determined with `flavour tagging' process. The kinematically similar mode $\Bs\to D_{s}^{-}\pi^{+}$ is used as a control channel for the determination of the decay time efficiency and flavour tagging performance. The fitted \CP observables are listed in Table \ref{tab:table2}. With the value of $\beta_{s}$ from previous LHCb measurment in $\BsjpsiKK$~\cite{Aaij:2014zsa} and $\Bs\to \jpsi \pi^{+}\pi^{-}$~\cite{Aaij:2014dka}, the angle $\gamma$ is determined as $\gamma=(127^{+17}_{-22})^{\circ}$.
\begin{table}[t]
\begin{center}
\begin{tabular}{l|c}  
Parameter&Value\\ \hline
$C_{f}$&$0.735\pm0.142\pm0.048$\\
$A_{f}^{\Delta\Gamma}$&$0.395\pm0.277\pm0.122$\\
$A_{\bar{f}}^{\Delta\Gamma}$&$0.314\pm0.274\pm0.107$\\
$S_{f}$&$-0.518\pm0.202\pm0.073$\\
$S_{\bar{f}}$&$-0.496\pm0.197\pm0.071$\\ \hline
\end{tabular}
\caption{ Results of the maximum likelihood fit to the selected $\Bs\to D^{\mp}_{s}K^{\pm}$ candidates. The first uncertainty is statistical and the second is systematic. }
\label{tab:table2}
\end{center}
\end{table}

\section{\CP violation searches in the $b$-baryons decays}
In the SM, decays of the \Lb bayron are predicted to have non-negligible \CP asymmetries, as large as $20\%$ for certain three-body decay modes~\cite{Hsiao:2014mua}. 
Large numbers of beauty baryons are produced at the LHC, the production ratio of $B^{0}:\Lb:\Bs$ particles is approximately $4:2:1$ within the LHCb detector acceptance~\cite{Aaij:2011jp}, which allows to perform sensitive studies of \CP violation in $b$-baryon sector.

\subsection*{\CP violation searches in $\Lb\to p\pi^{-}\pi^{+}\pi^{-}$}
The $\Lb\to p\pi^{-}\pi^{+}\pi^{-}$ decay is mediated by the weak interaction, \CP violation in decay can arise from the interference of a tree amplitude and a penguin amplitude.  

\begin{figure}[htb]
\centering
\includegraphics[width=0.40\textwidth]{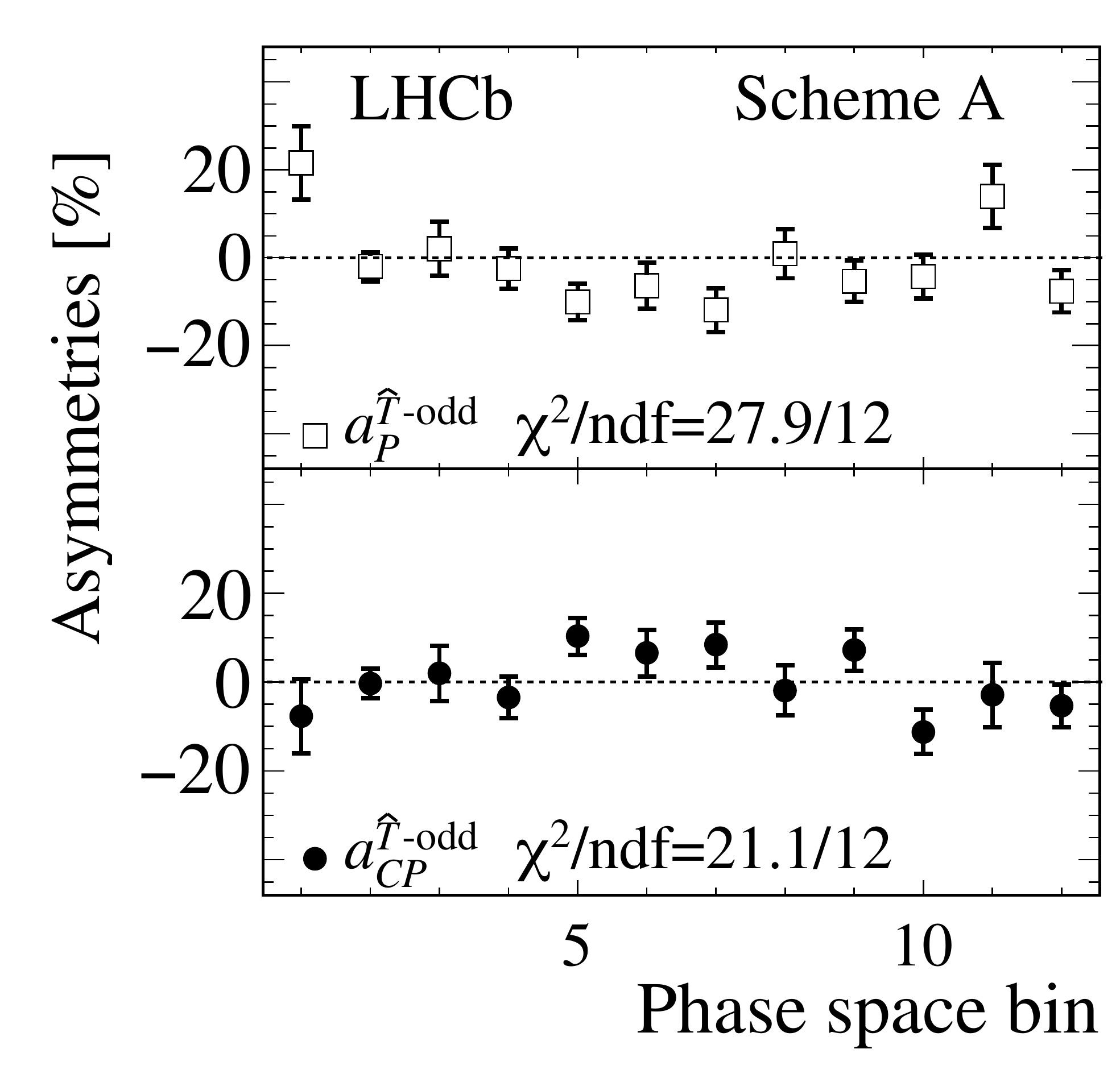}
\includegraphics[width=0.40\textwidth]{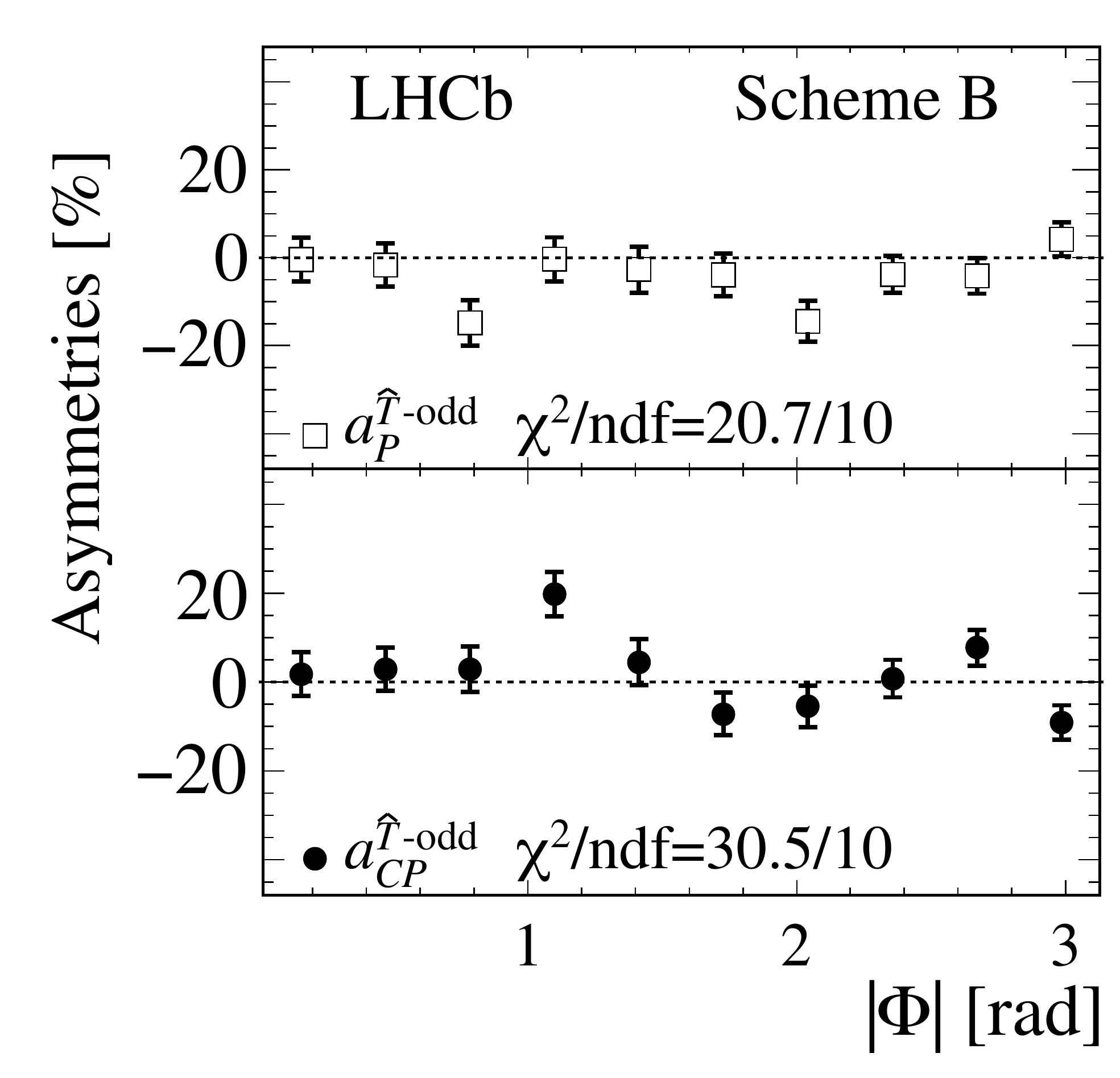}\\
\caption{ The asymmetries $a^{\hat{T}-odd}_{P}$ (upper panels) and $a^{\hat{T}-odd}_{\CP}$ (lower panels) for $\Lb\to p\pi^{-}\pi^{+}\pi{-}$ decays in bins of phase space and $\Phi$. }
\label{fig:figure1}
\end{figure}

The scalar triple products of final-state particle momenta in the \Lb centre-of-mass frame are used for $P$ and \CP violating searches, defined as $C_{\hat{T}}=\vv{p_{p}}\cdot(\vv{p}_{\pi^{-}_{\mathrm{slow}}}\times \vv{p}_{\pi^{+}})\propto \mathrm{sin}\Phi$ for \Lb, and  $\overline{C}_{\hat{T}}=\vv{p_{\bar{p}}}\cdot(\vv{p}_{\pi^{+}_{\mathrm{slow}}}\times \vv{p}_{\pi^{-}})\propto\mathrm{sin}\overline{\Phi}$ for $\overline{\varLambda}{}^{0}_{b}$, where $\Phi$ is the angle between the decay plane of the proton and higher-momentum $\pi^{-}$ and that of the $\pi^{+}$ and the lower-momentum $\pi^{-}$.
Two $\hat{T}$-odd asymmetries are built as
\begin{equation}
A_{\hat{T}}=\frac{N_{\Lb}(C_{\hat{T}}>0)-N_{\Lb}(C_{\hat{T}}<0)}{N_{\Lb}(C_{\hat{T}}>0)+N_{\Lb}(C_{\hat{T}}<0)},\ \overline{A}_{\hat{T}}=\frac{N_{\overline{\varLambda}{}^{0}_{b}}(-\overline{C}_{\hat{T}}>0)-N_{\overline{\varLambda}{}^{0}_{b}}(-\overline{C}_{\hat{T}}<0)}{N_{\overline{\varLambda}{}^{0}_{b}}(-\overline{C}_{\hat{T}}>0)+N_{\overline{\varLambda}{}^{0}_{b}}(-\overline{C}_{\hat{T}}<0)}.
\end{equation}
The $P$ and \CP violating observables can be constructed as
\begin{equation}
a_{P}^{\hat{T}-\mathrm{odd}}=\frac{1}{2}(A_{\hat{T}}+\overline{A}_{\hat{T}}),\ a_{\CP}^{\hat{T}-\mathrm{odd}}=\frac{1}{2}(A_{\hat{T}}-\overline{A}_{\hat{T}}).
\end{equation}
The analysis was performed in ten bins of $|\Phi|$ and in twelve bins of the phase space. The distribution of the asymmetries are shown in Figure \ref{fig:figure1}.
Combining the results of both schemes gives evidence for \CP violation in $\Lb\to p\pi^{-}\pi^{+}\pi^{-}$ at the level of  $3.3\sigma$~\cite{Aaij:2016cla}.

\subsection*{\CP violation searches in $\Lb\to pK^{-}\mu^{+}\mu^{-}$}
The $\Lb\to pK^{-}\mu^{+}\mu^{-}$ is mediated by flavour-changing neutral-current transitions in the SM. The theoretical prediction of \CP violation in this decay is very small~\cite{Paracha:2014hca,Alok:2011gv} which makes it particularly sensitive to \CP violation contributions from BSM physics.
The signal yield is determined to be $600\pm44$. 
The asymmetry $A_{\CP}$ is related to the raw asymmetry $A_{\mathrm{raw}}$ via
\begin{equation}
A_{\mathrm{raw}}\equiv\frac{N(\Lb\to pK^{-}\mu^{+}\mu^{-})-N(\overline{\varLambda}{}^{0}_{b}\to \bar{p} K^{+}\mu^{-}\mu^{+})}{N(\Lb\to pK^{-}\mu^{+}\mu^{-})+N(\overline{\varLambda}{}^{0}_{b}\to \bar{p} K^{+}\mu^{-}\mu^{+})}\approx A_{\CP}(\Lb\to pK^{-}\mu^{+}\mu^{-}) + A_{\mathrm{prob}}(\Lb)-A_{\mathrm{reco}}(K^{+})+A_{\mathrm{reco}}(p),
\end{equation}
where $A_{\mathrm{prob}}(\Lb)$ is the \Lb production asymmetry, $A_{\mathrm{reco}}(K^{+})$ and $A_{\mathrm{reco}}(p)$ are the reconstruction asymmetries for kaons and protons. 
Measurement of the difference of raw asymmetries between the signal and $\Lb\to pK^{-}\jpsi$ control mode, can cancel the effects from the production and reconstruction asymmetries. 
Thus two different \CP violating observables that are sensitive to different manifestations of \CP violation, $\Delta A_{\CP}$ and $a^{\hat{T}-odd}_{\CP}$, are measured. 
The \CP-odd observable is defined following the same method in the above $\Lb\to p\pi^{-}\pi^{+}\pi^{-}$ analysis. The results, $\Delta A_{\CP}=(-3.5\pm5.0(\mathrm{stat})\pm0.2(\mathrm{syst}))\times 10^{-2},\ a^{\hat{T}-odd}_{\CP}=(1.2\pm5.0(\mathrm{stat})\pm0.7(\mathrm{syst}))\times 10^{-2}$~\cite{Aaij:2017mib}, are comaptible with \CP and parity conservation and agree with SM predictions.

\section{Conclusions}
Many new measurements of \CP violation searches in $b$-hadron decays are performed by the LHCb experiment. These results improve our knowledge on the CKM parameters. All the measurements are still statistically limited and can be improved with LHC Run-II data.

\end{document}